\documentclass[aps,showpacs,preprintnumbers,amsmath, amssymb]{revtex4}

\usepackage{float}
\usepackage{graphics,epsfig}
\usepackage{graphicx}
\usepackage{dcolumn}
\usepackage{bm}

\begin{document}
\baselineskip=0.8 cm
\title{{\bf Holographic superconductor developed in BTZ black hole background with backreactions}}

\author{Yunqi Liu } \affiliation{Department of Physics, Fudan University, 200433 Shanghai, China}
\author{Qiyuan Pan, Bin Wang}   \affiliation{INPAC and Department of Physics, Shanghai Jiao Tong University, 200240 Shanghai, China}

\vspace*{0.2cm}
\begin{abstract}
\baselineskip=0.6 cm
\begin{center}
{\bf Abstract}
\end{center}

We develop a holographic superconductor in BTZ black hole background with  backreactions. We investigate the influence of the backreaction on the
condensation of the scalar hair and the dynamics of perturbation in the background spacetime. When the Breitenlohner-Freedman bound is
approached, we argue that only one of two possible operators can reflect the real property of the condensation in the holographic superconductor.
This argument is supported by the investigation in dynamics.

\end{abstract}

\maketitle
\newpage
\vspace*{0.2cm}

\section{Introduction}

Inspired by the anti-de Sitter/conformal field theories (AdS/CFT) correspondence, it has been argued that there exists remarkable connection
between the gravitational physics and the condensed matter physics (for reviews, see Refs. \cite{HartnollRev,HerzogRev,HorowitzRev}). It was
shown that when the temperature of the black hole drops below the critical value, the bulk AdS black hole becomes unstable and scalar hair
condenses on the black hole background. The emergence of the scalar hair in the bulk AdS black hole corresponds to the formation of a charged
condensation in the boundary dual CFTs. Due to the potential applications to the condensed matter physics, the condensation in the AdS black hole
background has been investigated extensively, see for example \cite{HorowitzPRD78}-\cite{JWChen}. A lot of studies have been done in the probe
approximation. Recently attempts to study the holographic superconductor away from the probe limit by considering the backreaction have been
carried out in \cite{HartnollJHEP2008}-\cite{Akhavan}.

Most holographic superconductors were constructed in (2+1)-dimensions or higher-dimensions. Recently the (1+1)-dimensional holographic
superconductor was constructed by using the $AdS_{3}/CFT_2$ correspondence \cite{ren}. The Banados-Teitelboim-Zanelli (BTZ) black hole was taken
as the gravity background in the construction. This background is interesting, since in this background the AdS/CFT correspondence was first
quantitatively proved by finding the quasinormal frequencies of the perturbation in the bulk spacetime in coincidence with the poles of the
correlation function in the dual CFT \cite{Bin1}. In \cite{ren} distinctive features in both normal and superconducting phases in the
(1+1)-dimensional holographic superconductor were disclosed in the probe limit. Since in the $AdS_3$ spacetime, the full current-current
correlation function is analytically solvable at finite temperature and zero chemical potential, the frequency and momentum dependence of the
conductivity  which describes the linear response to both temporally and spatially oscillating electric field was obtained. It is of interest to
generalize the study of the (1+1)-dimensional holographic superconductor away from the probe limit by considering the backreaction. This will be
the first task in our work. Furthermore we will study the dynamical properties of the perturbation of charged scalar field in the BTZ background,
which can help us further understand the constructed (1+1)-dimensional holographic superconductor.

In studying the holographic superconductor, one has to analyze the asymptotic behavior of a scalar field of mass $m$ in bulk AdS spacetime with
the form $\psi\sim \frac{\psi_-}{r^{\lambda-}}+\frac{\psi_+}{r^{\lambda+}}$, where $\lambda_\pm=(d\pm\sqrt{d^2+4 m^2l^2})/2$, $d$ is the
dimension of the space and $l$ the AdS radius. Usually, normalizability requires that the leading coefficient in $\psi$ must vanish. However,
since we have chosen a mass close to the Breitenlohner-Freedman (BF) bound, even the leading term in $\psi$ is normalizable. In this case, one
has a choice to consider solutions either with $\psi_- = 0$ or $\psi_+ = 0$. In the dual theory, the operator charged under the U(1) is dual to
$\psi$. When $m^2$ is close to the BF bound, there are two possible operators depending on how one quantizes $\psi$ in the bulk. Depending on the
choice of boundary conditions, we can read off the expectation value of an operator $\langle\mathcal{O}_{-} \rangle\sim\psi_-$, or of an operator
$\langle\mathcal{O}_{+} \rangle\sim\psi_+$ \cite{HartnollRev}. In order to apply the formalism in gravity to study the real condensed matter
physics, one may ask which one of the two possible operators can really reflect the properties of the real condensation. In \cite{Hart}, it was
found that both operators condensate qualitatively similar to that obtained in BCS theory as observed in many materials when the temperature
drops under a critical value, but one of them appears divergence at very low temperature. This divergence was expected to be cured by considering
the backreaction of the scalar field on the bulk metric. In \cite{Gregory,Pan-Wang}, it was argued that the condensation gap indicated in the
operator marks the ease of the scalar hair to be formed in the AdS black hole background. Whether the easiness of the formation of scalar hair
can be reflected consistently in two operators is a question to be asked. In the study of the Gauss-Bonnet effect on the condensation in the
probe limit, it was found that two operators cannot reflect the consistent behavior in the condensation influenced by the Gauss-Bonnet factor
except that we consider the direct signature of Gauss-Bonnet factor in the scalar mass by selecting the value $m^2l_{eff}^2$ instead of $m^2l^2$
\cite{Pan-Wang}. It is of interest to examine whether these two operators can reflect the consistent behavior of condensation when the
backreaction is taken into account. In this work, we will concentrate our attention on (1+1)-dimensional holographic superconductor and
generalize our discussion to higher dimensions.

The outline of this work is as follows. In section II, we discuss the fully backreacted one-dimensional holographic superconductor and
investigate the dependence of condensate on the backreaction. In section III, we study the dynamics of the backreacted BTZ black hole. We will
conclude in the last section of our main results.

\section{BTZ Holographic Superconductor}

The general action describing a charged, complex scalar field in the 3-dimensional Einstein-Maxwell action with negative cosmological constant
reads \cite{ren}
\begin{eqnarray}\label{action}
S=\frac{1}{2\kappa^2}\int d^3 x\sqrt{-g}(R+\frac{2}{l^2})+\int d^3
x\sqrt{-g}\left[-\frac{1}{4}F^{ab}F_{ab}-|\nabla\psi-i q A\psi
|^2-m^2|\psi|^2\right],
\end{eqnarray}
where $\kappa$ is the three dimensional gravitational constant $\kappa^2=8 \pi G_3$, and $G_3$ is the (2+1)-dimensional Newton constant,~$g$ is
the determinant of the metric,~$l$ is the AdS radius, $q$ and $m$ represent the charge and the mass of the scalar field respectively. In order to
consider the effect of the backreaction of the holographic superconductor, we take a metric ansatz as follows
\begin{eqnarray}\label{ansatz}
ds^2=-f(r)e^{-\chi(r)}dt^2+\frac{dr^2}{f(r)}+\frac{r^2}{l^2}dx^2~.
\end{eqnarray}
The electromagnetic field and the scalar field can be chosen as
\begin{eqnarray}
A_t=\phi(r)dt,~~~~\psi=\psi(r),
\end{eqnarray}
where $\psi(r)$ can be taken to be real without loss of generality. The Hawking temperature of this black hole, which will be interpreted as the
temperature of the CFT, can be expressed as
\begin{eqnarray}\label{temperature}
T=\left.\frac{f'(r)e^{-\chi(r)/2}}{4 \pi}\right|_{r=r_h}~.
\end{eqnarray}
Considering the ansatz of the metric, the equations of motion can be easily obtained
\begin{eqnarray}\label{equationsofmotion}
0&=&\psi ''(r)+\psi '(r)
\left[\frac{1}{r}+\frac{f'(r)}{f(r)}-\frac{\chi '(r)}{2}\right]+\psi
(r) \left[\frac{q^2 \phi (r)^2 e^{\chi
(r)}}{f(r)^2}-\frac{m^2}{f(r)}\right]~,\nonumber\\
0&=&\phi ''(r)+\phi '(r) \left[\frac{1}{r}+\frac{\chi '(r)}{2}\right]-\frac{2 q^2 \phi (r) \psi (r)^2}{f(r)}~,\nonumber\\
0&=&f'(r)+2 \kappa ^2 r \left[\frac{q^2 \phi (r)^2 \psi (r)^2
e^{\chi (r)}}{f(r)}+f(r) \psi '(r)^2+m^2 \psi (r)^2+\frac{1}{2}
 e^{\chi (r)} \phi '(r)^2\right]-\frac{2 r}{l^2},\nonumber\\
0&=&\chi '(r)+ 4 \kappa ^2 r \left[\frac{q^2 \phi (r)^2 \psi (r)^2
e^{\chi (r)}}{f(r)^2}+\psi '(r)^2\right],
\end{eqnarray}
where the prime denotes the derivative with respect to $r$. There are three useful scaling symmetries to be adopted in the above equations as in
\cite{Gregory}
\begin{equation}\label{symmetry1}
\phi \rightarrow \phi a,~\psi \rightarrow \psi a,~\kappa^2
\rightarrow \kappa^2 a^{-2},~q \rightarrow q a^{-1},
\end{equation}
\begin{equation}\label{symmetry2}
r\rightarrow r b, f\rightarrow f b^2,~\phi\rightarrow \phi b,
\end{equation}
\begin{equation}\label{symmetry3}
r\rightarrow rc,~l\rightarrow l c,~q\rightarrow q
c^{-1},~m\rightarrow m c^{-1}.
\end{equation}
We can use the symmetry (\ref{symmetry1}) to set $q=1$, and the
symmetries $(\ref{symmetry2}),(\ref{symmetry3})$ to set $r_h=1$ and
$l=1$.

We will use the shooting method to solve the equations of motion numerically with appropriate boundary conditions. There are two different
boundaries we need to consider. At the black hole horizon $r_h$ which is the root of $f(r_h)=0$, the solutions of the gauge and the scalar fields
have to be regular
\begin{eqnarray}\label{horizonboundry}
\phi(r_h)=0~,  ~~\psi'(r_h)=\frac{m^2}{f'(r_h)}\psi(r_h),
\end{eqnarray}
and the coefficients in the metric ansatz obey
\begin{eqnarray}\label{horizonmetric}
f'(r_h)&=&\frac{2 r_h}{l^2}-2 \kappa ^2 r_h \left[m^2 \psi
(r_h)^2+\frac{1}{2}
e^{\chi (r_h)} \phi '(r_h)^2\right],\nonumber\\
\chi '(r_h)&=&-4 \kappa ^2 r_h \left[\frac{q^2 \phi' (r_h)^2 \psi
(r_h)^2 e^{\chi (r_h)}}{f'(r_h)^2}+\psi '(r)^2\right].
\end{eqnarray}
At the spatial infinity, the asymptotic behaviors of the solutions are
\begin{eqnarray}
&&\chi\rightarrow 0~,~~~f(r)\sim \frac{r^2}{l^2}~,\nonumber\\
&&\phi(r)\sim \rho+ \mu \ln(r)~,~~~\psi(r)\sim
\frac{\psi_-}{r^{\lambda-}}+\frac{\psi_+}{r^{\lambda+}},
\end{eqnarray}
where $\lambda_\pm=1\pm\sqrt{1+l^2 m^2}$, $\mu$ is the chemical potential. When $\lambda_+$ $-$ $\lambda-$ $=2n$ $(n=1,2...)$, there appears a
logarithmic term  in the asymptotic behavior at infinity. For $-1\leq m^2<0$, both of these falloffs in $\psi(r)$ are normalizable, so one can
impose the boundary condition that either one vanishes. After imposing the condition that either $\psi_-$ or $\psi_+$ vanishes we have a one
parameter family of solutions. For $m^2 \geq0$, we can only relate the scalar operator in the field theory dual to the branch $\psi_+$ to
describe the condensation.

In order to find the effect of backreaction on the scalar
condensation, we need to count on numerical calculations.  In the
following we list our numerical results in solving Eq.
(\ref{equationsofmotion}) for different strength of the backreaction
$\kappa^2$ with $m^2=0$ and $m^2=-1$ respectively.

\begin{figure}[h]
\includegraphics[width=200pt]{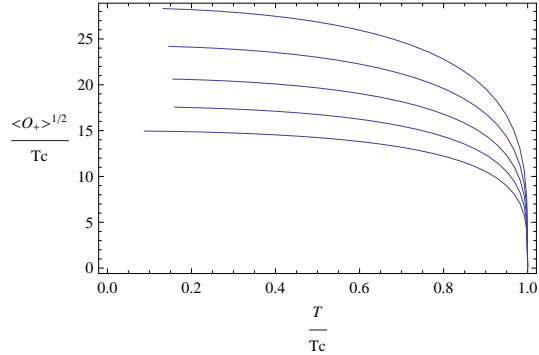}
\caption{\label{operatorplus0} Plot of condensate as a function of
temperature with $m^2=0$. Lines from bottom to top correspond to
$\kappa^2=0~,0.05~,0.1~,0.15$ and $0.2$ respectively.}
\end{figure}

When choosing the mass of the scalar field  $m^2=0$, the asymptotic
behavior of the scalar field at infinity takes the form
\begin{eqnarray}
\psi(r)=\psi_- +\frac{\psi_+}{r^{2}}~.
\end{eqnarray}
Only $\psi_+$ can be chosen dual to the scalar operator in the field theory to describe the condensate. In Table \ref{Table.I}, we list the
critical temperature for the operator starts to condense for different strength of the backreaction. We found that with the increase of the
backreaction, the critical temperature decreases. Thus the effect of the backreaction is to make it harder for scalar hair to form. This property
can also be seen from the condensation as shown in Fig. \ref{operatorplus0}. The condensation of the operator $\langle\mathcal{O}_{+} \rangle$
can start when  gap becomes higher for the stronger backreaction, which means that the scalar hair can be formed more difficult when the
backreaction is stronger. We can fit these data near the critical point and find that $\langle\mathcal{O}_{+} \rangle\sim(T_c-T)^{1/2}$ as
expected from mean field theory. The exponent $1/2$ implies that the phase transition is of the second order. The order of the phase transition
is not changed when the backreaction is taken into account. When the temperature tends to zero, the condensate tends to a finite value, which is
qualitatively similar to that observed in the BCS theory.

\begin{center}
\begin{table}[ht]
\caption{\label{Table.I} The dependence of the critical temperature
$T_{c}$ on the backreaction $\kappa^2$ for $m^2=0$. Obviously, the
critical temperature decreases as the backreaction grows.}
\begin{tabular}{|c|c|c|c|c|c|}
\hline
 $\kappa^{2}$ & 0  &  0.05  &  0.1  &  0.15  &  0.2 \\
[0.5ex] \hline
$T_{c}/\mu$ &~0.0460~&~0.0368~&~0.0295~&~0.0236~&~0.0189~\\
\hline
\end{tabular}
\end{table}
\end{center}

When the mass of the scalar field is chosen near the BF bound (we show for example $m^2=-1$ below), the asymptotic behavior of the scalar field
exhibits
\begin{eqnarray}
\psi(r)=\frac{\psi_-}{r} \ln(r)+\frac{\psi_+}{r}.
\end{eqnarray}
We have a choice to consider solutions either with  $\psi_+=0$ or $\psi_-=0$.


\begin{center}
\begin{table}[ht]
\caption{\label{Table.II} The dependence of the critical temperature
$T_{c}$ on the backreaction $\kappa^2$ for $m^2=-1$ with $\psi_-=0$.
Obviously, the critical temperature decreases as the backreaction
grows.}
\begin{tabular}{|c|c|c|c|c|c|}
\hline $\kappa^{2}$ & 0  &  0.05  &  0.1  &  0.15  &  0.2 \\ [0.5ex]
\hline
 $T_{c}/\mu$ &~0.136~&~0.133~&~0.131~&~0.128~&~0.126~\\ \hline
\end{tabular}
\end{table}
\end{center}

\begin{center}
\begin{table}[ht]
\caption{\label{Table.III} The dependence of the critical temperature $T_{c}$ on the backreaction $\kappa^2$ for $m^2=-1$ with $\psi_+=0$.
Obviously, the critical temperature decreases as the backreaction grows.}
\begin{tabular}{|c|c|c|c|c|c|}
\hline $\kappa^{2}$ & 0  &  0.05  &  0.1  &  0.15  &  0.2 \\ [0.5ex] \hline
 $T_{c}/\mu$ &~0.050~&~0.043~&~0.038~&~0.034~&~0.030~\\ \hline
\end{tabular}
\end{table}
\end{center}

 The dependances of the critical temperature on the backreaction for the operators start to condense in the field theory are shown in Tables
\ref{Table.II} and \ref{Table.III}  for choosing $\psi_-=0$ and $\psi_+=0$, respectively. Combining with  Table \ref{Table.I}, we find that no
matter how one quantizes $\psi$ in the bulk, the critical temperatures consistently drop when the backreaction grows, which shows that both
operators condense more difficult when the backreaction becomes stronger. This property will not be altered when we change the mass of the scalar
field.

In Fig. 2, we showed the condensations by plotting the expectation values of operators  $\langle\mathcal{O}_{-} \rangle\sim\psi_-$ and
$\langle\mathcal{O}_{+} \rangle\sim\psi_+$. We find that the operator $\langle\mathcal{O}_{-} \rangle$ can reflect the condensation when the
backreaction is taken into account. The gap of the condensation becomes higher when the backreaction becomes stronger, which indicates the
consistent picture shown in $T_c$ that the backreaction makes the condensation to be formed harder. When the BF bound is approached, the operator
$\langle\mathcal{O}_{+} \rangle$ shows drastically different behavior from that exhibited in Fig.1. In the probe approximation, the operator
appears divergent at very low temperature as observed in higher dimensions [25]. This divergence can be cured by considering the backreaction as
expected. Furthermore, when the BF bound is approached, the backreaction effect makes the gap of condensation lower in the operator
$\langle\mathcal{O}_{+} \rangle$, which shows different condensation behavior as illustrated in Fig. 1. This result disagrees with the behavior
exhibited in the critical temperature and cannot correctly  reflect the ease of the scalar hair to be formed in the AdS black hole background
when the backreaction is considered. This phenomenon does not only appear in (1+1)-dimensional superconductor. Actually different effects due to
the backreaction on condensations were also shown in two different operators in (2+1)-dimensional superconductors, see Fig. 1 in
\cite{HartnollJHEP2008}. When the BF bound is approached, one branch in the asymptotic behavior of the scalar field behaves as $\sim 1/r$ at the
spatial infinity. If we relate the expectation value of the operator to this branch, like $\langle\mathcal{O}_{+} \rangle$ above and
$\langle\mathcal{O}_{1} \rangle$ in \cite{HartnollJHEP2008}, in addition to the divergence appears at very low temperature in the operator in the
probe limit, the condensation gap indicated in the operator cannot reflect the correct extent of the scalar hair to be formed. This shows that
not both of the operators can reflect correctly of the condensation. We need to discard one of them when we are close to the BF bound. In the
following section, we will further show this result from the study of the dynamical properties.

\begin{figure}[h]
\includegraphics[width=200pt]{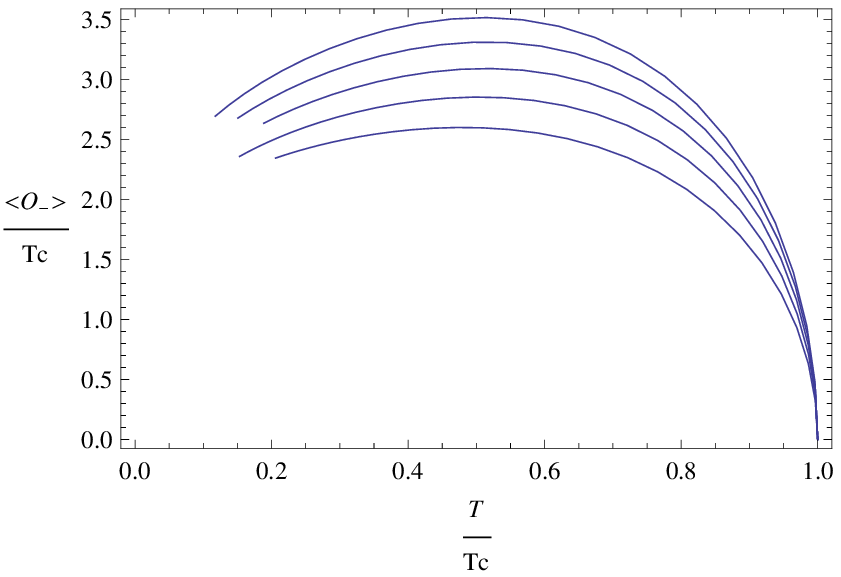}
\includegraphics[width=200pt]{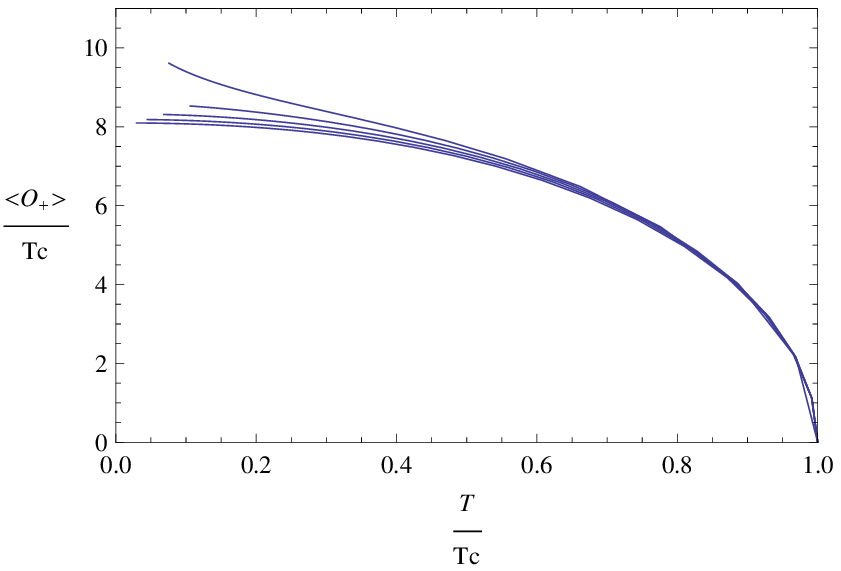}
\caption{\label{operatorminus-1}  Plot of the condensates as  functions of temperature for $m^2=-1$ with $\psi_+=0$ (the left panel) and
$\psi_-=0$ (the right panel), respectively. In the left panel five lines from bottom to top correspond to $\kappa^2=0~,0.05~,0.1~,0.15~,0.2$
respectively, but lines in the right panel show different order of strength of the backreaction. }
\end{figure}

\section{Dynamics of the backreacted BTZ Black Hole}

In this part, we begin to investigate the dynamics of a backreacted BTZ black hole when the black hole approaches marginally stable from the high
temperature. We will consider the minimally coupled, charged scalar perturbation, $\psi_{\varpi,~q}(r)e^{-i(\omega t+kx)}$, with mass $m$ obeying
the wave equation
\begin{eqnarray}\label{simplemotion}
0&=&\psi_{\varpi,\mathfrak{q}} ''(r)+\psi_{\varpi,\mathfrak{q}} '(r)
\left[\frac{1}{r}+\frac{f'(r)}{f(r)}-\frac{\chi
'(r)}{2}\right]+\psi_{\varpi,\mathfrak{q}} (r) \left\{\frac{[\omega
+q \phi(r)]^2 e^{\chi
(r)}}{f(r)^2}-\frac{m^2}{f(r)}-\frac{l^2 k^2}{f(r) r^2}\right\}~,\\
0&=&\phi ''(r)+\phi '(r) \left[\frac{1}{r}+\frac{\chi '(r)}{2}\right],\\
0&=&f'(r)+\kappa ^2 e^{\chi (r)} r \phi '(r)^2-\frac{2 r}{l^2},\\
0&=&\chi '(r),
\end{eqnarray}
where the prime denotes derivative with respect to $r$. The analytical solutions of the above equations are expressed as
\begin{eqnarray}\label{solution}
\chi(r)=C, ~~\phi(r)=\mu \ln\left(\frac{r}{r_h}\right),~~f(r)=-\kappa^2 e^{\chi(r)} \mu^2 \ln\left(\frac{r}{r_h}\right)+\frac{r^2-r_h^2}{l^2},
\end{eqnarray}
where $C$ is an integral constant.

In fact, we can use three similar scaling symmetries as expressed in
Eqs. (\ref{symmetry1})$-$(\ref{symmetry3}) to set $q=1$, $r_h=1$ and
$l=1$ as well. The difference is that we have to use the following
symmetry to replace (\ref{symmetry2})
\begin{equation}\label{newsymmetry2}
r\rightarrow r b, f\rightarrow f b^2, \phi\rightarrow\phi b,\omega\rightarrow\omega b^{-1},k\rightarrow k b^{-1}.
\end{equation}
Eq. (\ref{simplemotion}) can be rewitten as
\begin{equation}\label{perturbation equation}
0=\psi_{\varpi,\mathfrak{q}} ''(u)+\psi_{\varpi,\mathfrak{q}} '(u)
\left[\frac{1}{u}+\frac{g'(u)}{g(u)}\right]+\psi_{\varpi,\mathfrak{q}}
(u) \left[\frac{(\varpi +
\aleph(u))^2}{g(u)^2}-\frac{m^2}{g(u)}-\frac{l^2
\mathfrak{q}^2}{g(u) u^2}\right],
\end{equation}
with
\begin{eqnarray}
g(u)&=&-\kappa^2 \sigma^2 \ln(u)+\frac{u^2-1}{l^2},\nonumber\\
\sigma&=&\frac{\mu}{r_h}e^{\frac{C}{2}},\nonumber\\
\varpi&=&\frac{\omega}{r_h}e^{\frac{C}{2}},\nonumber\\
\mathfrak{q}&=&\frac{k}{r_h},\nonumber\\
\aleph(u)&=&q \sigma \ln(u) ,
\end{eqnarray}
where $u=r/r_h$ and the prime here denotes the derivative with respect to $u$. Now $u=\infty$ is the AdS boundary and $u=1$ is the location of
the horizon. The Hawking temperature of the black hole is
\begin{eqnarray}\label{hawkingtemperature}
T=\frac{2-\kappa^2 \sigma^2}{4 \pi \sigma} \mu.
\end{eqnarray}

Near the AdS boundary $u\sim \infty$, Eq. (\ref{perturbation
equation}) becomes
\begin{equation}
\psi_{\varpi,\mathfrak{q}} ''(u)+\frac{3}{u}
\psi_{\varpi,\mathfrak{q}} '(u)- \frac{
m^2}{u^2}\psi_{\varpi,\mathfrak{q}} (u)=0~,
\end{equation}
and $\psi_{\varpi,\mathfrak{q}}(u)$ has a fall-off behavior as
\begin{equation}
\psi_{\varpi,\mathfrak{q}}(u)\sim\frac{
\psi_{\varpi,\mathfrak{q}}^{-}}{u^{\lambda_{-}}}+\frac{\psi_{\varpi,\mathfrak{q}}^{+}}{u^{\lambda_{+}}},
\end{equation}
where $\lambda_{\pm}=1\pm\sqrt{1+m^{2}l^{2}}$ are characteristic exponents of the perturbation equation.

Following the AdS/CFT dictionary, the order parameter expectation
value $\langle\mathcal{O}_{\varpi,\mathfrak{q}}\rangle$ corresponds
to $\psi_{\varpi,\mathfrak{q}}^{+}$ while the source term is
$\psi_{\varpi,\mathfrak{q}}^{-}$, so the response function can be
defined by~\cite{maeda}
\begin{equation}\label{response function}
\chi_{\varpi,\mathfrak{q}}:=\left.\frac{\delta\langle\mathcal{O}_{\varpi,\mathfrak{q}}\rangle}{\delta\psi_{\varpi,\mathfrak{q}}^{-}}\right
|_{\psi_{\varpi,\mathfrak{q}}^{-}\rightarrow0}\propto\frac{\psi_{\varpi,\mathfrak{q}}^{+}}{\psi_{\varpi,\mathfrak{q}}^{-}}~.
\end{equation}
Also we can choose the other way round, the expectation value $\langle\mathcal{O}_{\varpi,\mathfrak{q}}\rangle$ corresponds to
$\psi_{\varpi,\mathfrak{q}}^{-}$ while the source term is $\psi_{\varpi,\mathfrak{q}}^{+}$, so that
$\chi_{\varpi,\mathfrak{q}}\propto\frac{\psi_{\varpi,\mathfrak{q}}^{-}}{\psi_{\varpi,\mathfrak{q}}^{+}}$. Since we aim to investigate the
perturbation, we have to solve the equation of motion of the scalar field, Eq. (\ref{perturbation equation}), based on the boundary conditions at
the horizon and at the AdS boundary. After obtaining the coefficients $\psi_{\varpi,\mathfrak{q}}^{\pm}$, we can discuss the behavior of the
response function.

Near the horizon $u\sim1$, we should impose the incoming wave
boundary condition
\begin{equation}\label{incoming}
\psi_{\varpi,~q}(u)\sim(u-1)^{- i\frac{\omega}{4\pi T}}.
\end{equation}
Introducing a new variable $\varphi$ as
$\psi_{\varpi,\mathfrak{q}}(u)=\mathcal{\Re}(u)\varphi_{\varpi,\mathfrak{q}}(u)$
and choosing
$\mathcal{\Re}(u)=exp[-i\int^{u}_{1}\frac{\varpi+\aleph(u)}{g(u)}]$,~which
asymptotically approaches  Eq. (\ref{incoming}) at the horizon, one
can express the boundary condition at the horizon as
$\left.\varphi_{\varpi,~q}\right|_{u=1}=const.$, and Eq.
(\ref{perturbation equation}) becomes
\begin{eqnarray} \label{main equation}
0&=&\varphi_{\varpi,\mathfrak{q}} ''(u)+B_1(u)
\varphi_{\varpi,\mathfrak{q}} '(u)+B_2(u)
\varphi_{\varpi,\mathfrak{q}}(u)~,
\end{eqnarray}
with
\begin{eqnarray}
B_1(u)&=&\frac{g'(u)}{g(u)}+\frac{2 i[\aleph (u)+\varpi]}{
g(u)}+\frac{1}{u}~,\nonumber\\
B_2(u)&=&-\frac{m^2}{g(u)}-\frac{l^2 \mathfrak{q}^2}{u^2
g(u)}-\frac{i [\aleph (u)+\varpi+u \aleph'(u) ]}{g(u)}~.
\end{eqnarray}

Near the AdS boundary $u\sim \infty$,
$\varphi_{\varpi,\mathfrak{q}}$ behaves as
\begin{equation}\label{asymptotic behavior}
\varphi_{\varpi,\mathfrak{q}}(u)\sim
\frac{\varphi_{\varpi,\mathfrak{q}}^{-}}{u^{\lambda_{-}}}+\frac{\varphi_{\varpi,\mathfrak{q}}^{+}}{u^{\lambda_{+}}}.
\end{equation}
The boundary conditions at the horizon are now given by
\begin{eqnarray}\label{boundary}
\varphi_{\varpi,\mathfrak{q}}|_{~u=1}&=&1~,\nonumber\\
\left.\frac{\varphi^{\prime}_{\varpi,\mathfrak{q}}}{\varphi_{\varpi,\mathfrak{q}}}\right|_{~u=1}
&=&-\left.\frac{B_2(u)}{B_1(u)}\right|_{~u=1}~.
\end{eqnarray}
Eq. (\ref{main equation}) is a linear equation and
$\varphi_{\varpi,~q}(u)$ must be regular at the horizon. Since we do
not concentrate on the amplitude of
$\varphi_{\varpi,\mathfrak{q}}(u)$, we can set
$\left.\varphi_{\varpi,\mathfrak{q}}\right|_{u=1}=1$.

Eq. (\ref{main equation}) has to be solved numerically under the boundary conditions (\ref{boundary}). We will examine the behavior of the
charged scalar field perturbation which can present us an objective picture on how the black hole approaches the marginally stable mode when the
temperature drops and the backreaction becomes stronger.  Without loss of generality, hereafter we will consider $m^2l^2=-1$ as an explicit
example in our calculation.

The dimensionless parameter $\sigma$ determines the phase structure and its critical value $\sigma_{c}$ can be calculated numerically. When
$\kappa^2=0$, the blackground configuration goes back to the probe limit and the corresponding critical temperature $T_c$ should be consistent
with the result obtained  in \cite{ren}.

Taking $m^2l^2=-1$, near the AdS boundary,~$\varphi_{\varpi,\mathfrak{q}}(u)$ behaves as
\begin{equation}
\varphi_{0,~0}(u)\sim\frac{\varphi_{0,~0}^{-} \ln(u)}{u} +\frac{\varphi_{0,~0}^{+}}{u}.
\end{equation}
Choosing $\varphi_{0,~0}^{-}=0$, we get the critical point $\sigma_{c}$ and the critical temperature $T_{c}$ for $m^2=-1$ with different values
of the backreaction $\kappa^2$, i.e., $\kappa^2=0~,0.05~,0.1~,0.15$ and $0.2$, which have been presented in Table \ref{Table.III}. It is clear
that the results in Table \ref{Table.IV} are consistent with those in Table \ref{Table.II}.

\begin{center}
\begin{table}[ht]
\caption{\label{Table.IV} The dependence of the critical point $\sigma_{c}$ and the critical temperature $T_{c}$ on the backreaction $\kappa^2$
with $m^2=-1$. Obviously, $\sigma_{c}$ and $T_{c}$ decrease as the backreaction increases.}
\begin{tabular}{|c|c|c|c|c|c|}
\hline $\kappa^{2}$ & 0  &  0.05  &  0.1  &  0.15  &  0.2 \\ [0.5ex]
\hline
 $\sigma_{c}/\mu$ &~3.165~&~2.903~&~2.688~&~2.508~&~2.357~\\ [0.5ex]
\hline
 $T_{c}/\mu$ &~0.050~&~0.043~&~0.038~&~0.034~&~0.030~\\ \hline
\end{tabular}
\end{table}
\end{center}

\begin{figure}[h]
\includegraphics[width=180pt]{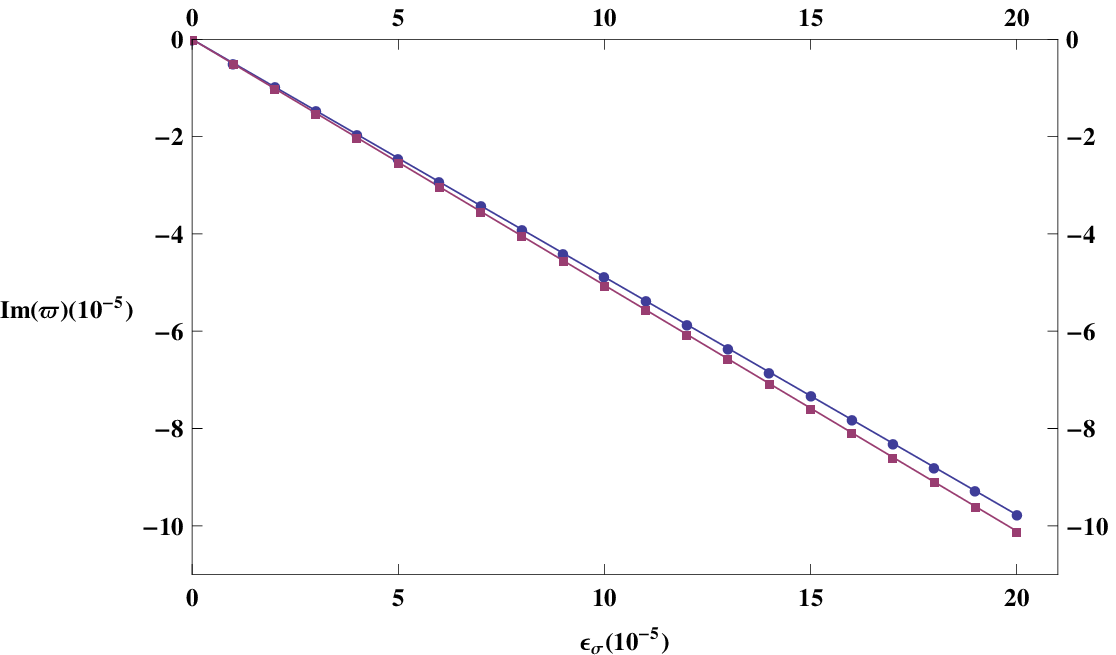}
\includegraphics[width=180pt]{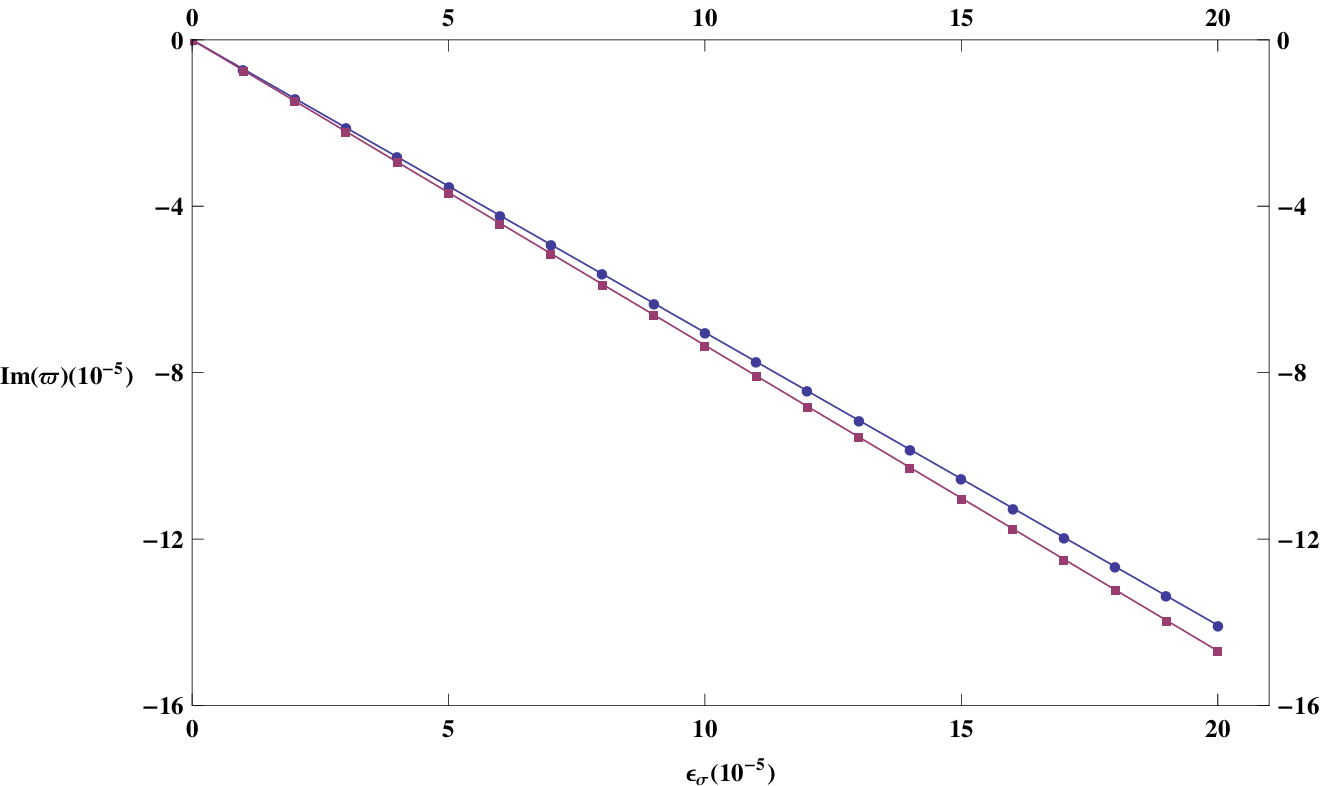}
\caption{\label{QNM} (Color online)  The trajectories of the imaginary parts of the lowest quasinormal frequency for different values of
$\kappa^2$ with $\varphi_{\varpi,~q=0}^{-}=0$ (the left panel) and $\varphi_{\varpi,~q=0}^{+}=0$ (the right panel), respectively. The blue line
corresponds to $\kappa^2=0$ while the pink one is $\kappa^2=0.2$. While the temperature drops to the critical point, the system approaches the
marginally stable mode.}
\end{figure}

Now we report the influence of the backreaction on the scalar
perturbation behavior. We concentrate on the lowest quasinormal
frequency which gives the relaxation time \cite{wang,liupanwang}. We
can obtain the quasinormal frequencies by solving Eq. (\ref{main
equation}) based on the boundary conditions (\ref{boundary}) at the
horizon and $\varphi_{\varpi,~q=0}^{-}=0$ at the AdS boundary. We
deviate $\sigma$ away from the critical value $\sigma_c$ and denote
the deviation by $\varepsilon_{\sigma}$,
$\varepsilon_{\sigma}=1-\sigma/\sigma_{c}=10^{-5}
n,~n=1,2,\cdots,20$. We see that all the imaginary parts of the
quasinormal frequencies are negative, which shows that the black
hole spacetime is stable. For the larger backreaction effect, the
imaginary part of the lowest quasinormal frequency has larger
deviation from zero. This implies that the stronger backreaction can
ensure the system to be more stable and can slow down the process to
make the high temperature black hole phase become marginally stable.
With the decrease of the black hole temperature, we see that the
lowest quasinormal frequency approaches the origin and vanishes when
the temperature of the system reaches the critical value, which
indicates that the system approaches marginally stable. We show the
object picture in Fig. \ref{QNM}. The lowest quasinormal frequency
approaches the origin with equal spacing and we fit the results for
different strength of the backreaction in polynomials as below
\begin{eqnarray}\label{QNMs}
\kappa^2&=&0,~~~~~~\varpi_{QNM} \sim (1.58-1.19i)\times10^{-13}
+(0.803-0.489i)~\varepsilon_{\sigma} -
(0.48+0.29i)~\varepsilon_{\sigma}^{2},\nonumber\\
\kappa^2&=&0.05,~~\varpi_{QNM} \sim (3.60-2.25i)\times10^{-12} +
(0.791-0.493i)~\varepsilon_{\sigma} +
(0.50-0.89i)~\varepsilon_{\sigma}^{2},\nonumber\\
\kappa^2&=&0.1,~~~~\varpi_{QNM} \sim (6.05-0.18i)\times10^{-15} +
(0.781-0.497i)~\varepsilon_{\sigma} -
(0.46+0.30i)~\varepsilon_{\sigma}^{2},\nonumber\\
\kappa^2&=&0.15,~~\varpi_{QNM} \sim (-1.47+0.95i)\times10^{-12} +
(0.770-0.501i)~\varepsilon_{\sigma} -
(0.45+0.30i)~\varepsilon_{\sigma}^{2},\nonumber\\
\kappa^2&=&0.2,~~~~\varpi_{QNM} \sim (3.89-2.62i)\times10^{-13} +(0.760-0.505i)~\varepsilon_{\sigma} - (0.44+0.31i)~\varepsilon_{\sigma}^{2}.
\end{eqnarray}
The effect of the backreaction on the scalar perturbation behavior
will not be altered when we set $\varphi_{\varpi,~q=0}^{+}=0$ at the
AdS boundary, see the right panel in Fig. \ref{QNM}.

Now we start to investigate the correlation length of the system. The critical behavior of the system near the critical point is determined by
the large-scale fluctuations. The correlation length is the scale parameter that exists in the system near the phase transition point. It
increases while the temperature approaches its critical value and becomes infinite at the moment of the phase transition. Now we check the
influence imposed by the backreaction on the correlation length $\xi$, which is defined by $\xi^{2}:=-\mathfrak{q}^{-2}$ \cite{maeda,liupanwang}.
We consider a perturbation with $\varpi=0$ for different $\kappa^2$, and solve Eq. (\ref{main equation}) with $\varpi=0$ under boundary
conditions: Eq. (\ref{boundary}) at the horizon and $\varphi_{\varpi=0,~q}^{+}=0$ at the AdS boundary first. The results for the correlation
length $\xi$ can be fitted by polynomials as
\begin{eqnarray}\label{xi}
\kappa^2&=&0.1,~~~~\xi^{-2} \sim 3.07\times10^{-11} + 9.08~\varepsilon_{\sigma},\nonumber\\
\kappa^2&=&0.15,~~\xi^{-2} \sim 6.84\times10^{-12} + 9.13~\varepsilon_{\sigma},\nonumber\\
\kappa^2&=&0.2,~~~~\xi^{-2} \sim 1.44\times10^{-11} + 9.24~\varepsilon_{\sigma}.
\end{eqnarray}
It is obvious that the correlation length $\xi$ depends on the backreacting parameter $\kappa^2$. For the smaller backreaction, the correlation
length is bigger for the same deviation from the critical point of the system, which means that it is easier for the system to approach the phase
transition point when the backreaction is smaller. This result is consistent with the influence of the backreaction on the critical temperature
and agrees to the property exhibited in the perturbation behavior.

When we set $\varphi_{\varpi,~q=0}^{-}=0$ at the AdS boundary, the condensation is expressed by the operator $\langle\mathcal{O}_{+} \rangle$.
The results for the correlation length $\xi$ are fitted by polynomials as listed below
\begin{eqnarray}\label{xi}
\kappa^2&=&0.1,~~~~\xi^{-2} \sim 4.27\times10^{-13} + 2.374~\varepsilon_{\sigma},\nonumber\\
\kappa^2&=&0.15,~~\xi^{-2} \sim 4.12\times10^{-12} + 2.360~\varepsilon_{\sigma},\nonumber\\
\kappa^2&=&0.2,~~~~\xi^{-2} \sim 1.59\times10^{-12} + 2.347~\varepsilon_{\sigma}.
\end{eqnarray}
This result implies that for the same deviation from the critical point of the system the correlation length is bigger when the backreaction is
bigger, which means that it is easier for the system to approach the phase transition point when the backreaction is stronger. This is in
contrary to the influence of the backreaction disclosed in the lowest quasinormal frequency and the critical temperature, which supports the
argument in the study of condensation that  $\langle\mathcal{O}_{+} \rangle$ is not an appropriate operator.

\section{Conclusions and discussions}

We developed the holographic superconductors in the BTZ black hole
background with backreactions. From the critical temperature and the
dynamical perturbation properties, we observed that the stronger
backreaction makes it more difficult for the scalar hair to
condensate.

In the holographic superconductor, we have two operators to describe
the condensation. It is of interest to ask which one can really
reflect the properties of the condensation. In the constructed
(1+1)-dimensional holographic superconductor, we observed that the
condensation read from expectation values of different operators
reflects different influence of the backreaction on the formation of
the scalar hair. This phenomenon was also observed in the study of
the effect of the backreaction in the (2+1)-dimensional holographic
superconductor \cite{HartnollJHEP2008}.

The operator $\langle\mathcal{O}_{-} \rangle$  in our paper is similar to $\langle\mathcal{O}_{2} \rangle$ [25].
The gaps of condensation increase with the increase of the
backreaction for these two operators as reported in our manuscript and in [25].
$\langle\mathcal{O}_{+} \rangle$  in our paper is similar to $\langle\mathcal{O}_{1} \rangle$ 
in [25], where these two gaps of the condensation increase when the
backreaction decreases.

In [25], $\langle\mathcal{O}_{1} \rangle$ is of mass dimension one operator, which
is related to the asymptotic behavior $\frac{\psi^{(1)}}{r}$ at infinity.
In that case, the mass was not equal to the BF bound.
In our case, when the BF bound is approached, $\langle\mathcal{O}_{+} \rangle$ can have the similar asymptotic behavior $\frac{\psi+}{r}$.

The condensation gap indicated in the operator marks the ease of the scalar hair to be formed[19,20]. It should reflect the consistent influence
of the backreaction as that shown in the critical temperature and dynamical perturbation. However, we found that one of the two operators cannot exhibit the consistency. We argued that this inconsistency tells us that this operator is not appropriate to describe
the condensation. We observed that this operator is usually associated with the branch with asymptotic behavior of the scalar field $\sim 1/r$ at
the spatial infinity. This argument is supported by the property of the correlation length disclosed in studying
the dynamics of the system.

\begin{acknowledgments}

This work was partially supported by the National Natural Science
Foundation of China.

\end{acknowledgments}

\end{document}